\documentclass[12pt]{article}
\pdfoutput=1
\usepackage{latexsym}
\usepackage[nosort]{cite}
\usepackage{epsfig,amssymb,euscript,slashed}
\usepackage[normalem]{ulem} 
\usepackage{amsmath}
\usepackage{float}
\usepackage[linktocpage]{hyperref}
\usepackage{bbm}
\usepackage{braket}
\usepackage{xcolor}
\usepackage{fancybox}
\usepackage{geometry}
\def\be{\begin{equation}}
\def\ee{\end{equation}}
\def\bseq{\begin{subequations}}
\def\eseq{\end{subequations}}

\def\bea{\begin{eqnarray}}
\def\eea{\end{eqnarray}}

\def\bseq{\begin{subequations}}
\def\eseq{\end{subequations}}

\numberwithin{equation}{section} 
\usepackage[]{graphicx}

\def\tr           {\mathop{\rm Tr}}

\def\sqr#1#2{{\vcenter{\vbox{\hrule height.#2pt
 \hbox{\vrule width.#2pt height#1pt \kern#1pt \vrule width.#2pt}\hrule
 height.#2pt}}}}

\def\slashchar#1{\setbox0=\hbox{$#1$}           
\dimen0=\wd0                                 
\setbox1=\hbox{/} \dimen1=\wd1               
\ifdim\dimen0>\dimen1                        
\rlap{\hbox to \dimen0{\hfil/\hfil}}      
#1                                        
\else                                        
\rlap{\hbox to \dimen1{\hfil$#1$\hfil}}   
/                                         
\fi}

\begin{document}
\font\cmss=cmss10 \font\cmsss=cmss10 at 7pt

\title{
 Tensionless/strong-field
 limit of M5-brane and conformal symmetry of its bosonic body
\\[0.5cm]
}

\author{Igor Bandos${}^{a,b}$, Kurt Lechner${}^{c,d}$ and Dmitri P. Sorokin${}^{a,c,d}$}

\date{August 19, 2026. 2607.29653 V2 [hep-th] }

\maketitle

\begin{center}

${}^a${\it
Department of Physics and EHU Quantum Center, University of the Basque Country UPV/EHU, \\
P.O. Box 644, 48080 Bilbao, Spain,

\vspace{0,3cm}
${}^b$ IKERBASQUE, Basque Foundation for Science,
48011, Bilbao, Spain}

\vspace{0,3cm}
${}^c${\it
Dipartimento di Fisica e Astronomia “Galileo Galilei”,
Universit\`a degli Studi di Padova,\\
Via Marzolo 8, 35131 Padova, Italy}

\vspace{0,3cm}
${}^d${\it
INFN, Sezione di Padova, Via Marzolo 8, 35131 Padova, Italy
}
\end{center}

\vspace{1cm}

\abstract{We obtain a tensionless limit of the M-theory 5-brane action and discuss its properties.
The consistency of this limit requires that the field strength $H_3$ of the 2-form gauge field on the M5 brane worldvolume is non-vanishing and non-degenerate, so that one can also regard it as a strong $H_3$-field limit.
 The induced $6d$ worldvolume metric of this  theory is non-degenerate in contrast to conventional tensionless (null) $p$-branes. We find that the tensionless M5-brane action  is invariant under a super-Weyl symmetry acting on the bulk supergeometry, while in the purely bosonic case the action is invariant under an 11D target-space conformal symmetry, a property which is characteristic for the tensionless bosonic branes. Upon imposing a static gauge, which fixes worldvolume diffeomorphisms, this 11D conformal symmetry becomes a deformed 6d worldvolume conformal symmetry whose action on the worldvolume coordinates involves field dependent terms. When the transversal fluctuations of the 5-brane in the bulk are zero, the model reduces to the conformal non-linear chiral 2-form electrodynamics considered earlier in \cite{Gibbons:2000ck,Townsend:2019ils}.
}
\thispagestyle{empty}

\newpage

\tableofcontents


\section{Introduction}

The main goal of this paper is to study properties of the model which arises by taking a tensionless limit of the M-theory 5-brane. In  \cite{Gibbons:2000ck} the M5-brane tensionless limit was considered in the on-shell superembedding formulation of the theory \cite{Howe:1996yn,Howe:1997fb}, but only for a simplified situation in which scalar and fermionic fluctuations of the M5-brane are set to zero, so that only a chiral 2-form gauge field propagates in the M5 six-dimensional worldvolume.
More recently in \cite{Townsend:2019ils} this limit was studied in the Hamiltonian formulation \cite{Bergshoeff:1998vx}  for the case of a planar and static 5-brane. In \cite{Gibbons:2000ck,Townsend:2019ils}  it was shown that in this case the limit of the zero M5-brane tension is equivalent to the strong worldvolume gauge field limit and that the resulting theory is a non-linearly interacting chiral 2-form electrodynamics in $6d$  which is conformally invariant and has an infinite number of conserved charges. Upon a dimensional reduction to $D=4$ this theory results in the conformally invariant non-linear electrodynamics of \cite{BialynickiBirula:1984tx,BialynickiBirula:1992qj} which was revisited in \cite{Mezincescu:2019vxk} also with regards to its relation to a tensionless D3-brane (whose action in a type IIB $D=10$ superbackground we present in Appendix \ref{tensionlessD3}).

Here we show that a complete action for a supersymmetric tensionless M5-brane can be obtained by taking a zero-tension limit of the original super-M5-brane action  \cite{Bandos:1997ui,Aganagic:1997zq}.
The $\kappa$-symmetry of the M5-brane is preserved in this limit, which indicates that the ground state of this dynamical system is a supersymmetric BPS state. However, somewhat unexpectedly, the chiral two-form becomes a kappa-symmetry singlet. This breaks the balance between fermionic and bosonic worldvolume physical degrees of freedom transformed under $\kappa$-symmetry and thus raises an issue of better understanding supersymmetry properties of this model. Another peculiarity of this limit is that the 6d conformal invariance of the pure chiral two-form Bialynicki--Birula--like theory is still preserved when bosonic fluctuations of the M5-brane are switch on, though conformal transformations of $6d$ worldvolume coordinates become field dependent. This 6d conformal symmetry arises in a static gauge (of worldvolume diffeomorphisms) and originates from an $11D$ target-space conformal invariance.  However, when fermionic fluctuations of the M5-brane are switched on, the $11D$ conformal symmetry gets broken rather than being extended to an $11D$ superconformal invariance. We leave the resolution of this `superconformal issue' to a future work, while in this paper we will discuss in detail conformal properties of the purely bosonic tensionless M5-brane.

We use the following conventions and notation.
The $6d$ worldvolume coordinates are denoted by $\xi^m$ $(m=0,1,2,3,4,5)$. The supercoordinates of the $D=11$ target superspace are denoted by $Z^{\underline{M}}=(X^{\underline m},\Theta^{\underline\mu})$, where $X^{\underline m}$ $(\underline m=0,1,\ldots, 10)$ are bosonic coordinates and $\Theta^{\underline\mu}$ $(\underline\mu=1,\ldots, 32)$ are fermionic ones. The lower indices from the beginning of the Latin alphabet stand for flat bulk tangent-space indices $({\underline a},{\underline b},,..=0,1,\ldots, 10)$ .
The space-time (and worldvolume) signatures are chosen to be mostly minus,
\be\eta_{\underline a\underline b}=\text{diag} (+1,-1,\ldots,-1)\; . \qquad  \ee

\section{M5-brane action}

In a generic 11D supergravity background the original $\kappa$-symmetric M5-brane action   \cite{Bandos:1997ui,Aganagic:1997zq}  is \footnote{A different variant of the action for the M5-brane was constructed in \cite{Ko:2013dka}. It is tailored to describe, in a simpler manner, M5-branes wrapping an internal 3d manifold and study their relation to the Bagger-Lambert-Gustavsson-like construction \cite{Bagger:2006sk,Gustavsson:2007vu,Bagger:2007jr,Bagger:2012jb} of multiple membranes with a gauge group of $3d$ volume preserving diffeomorphisms \cite{Ho:2008nn,Bandos:2008fr,Bandos:2008jv,Pasti:2009xc}.}
\begin{eqnarray} \label{SM5=}
S_{M5} &=&  - T \int_{\mathcal W^6} d^6\xi \sqrt{-det(g_{mn} + \frac 1 {\sqrt{T}}
\tilde{H}_{mn})}
- \frac 1 2 \int_{\mathcal W^6} \, v \wedge H_3 \wedge i_vH_3 + \nonumber \\
&& \qquad + \frac {T}{2} \int_{\mathcal W^6} ({A}_6 + \frac 1 {\sqrt{T}} db_2 \wedge
{A}_3) \; , \qquad
\end{eqnarray}
where $\xi^m$ ($m=0,1,2,3,4,5$) parametrize the $6d$ worldvolume $\mathcal W^6$, $T$ is the M5-brane tension of mass dimension $m^6$. \footnote{In \cite{Bandos:1997ui,Aganagic:1997zq} the M5-brane tension was set to one. Since we are interested in the tensionless limit, here we restore the explicit dependence  on the tension requiring the two-form gauge field $b_2$ be dimensionless (i.e. $b_{mn}$ has canonical mass dimension $m^2$ (see also \cite{Buratti:2019guq})).}  $A_3(Z(\xi))$ and $A_6(Z(\xi))$ are pullbacks on $\mathcal W^6$ of the $11D$ three-form gauge superfield and its dual six-form superfield in an $11D$ supergravity superspace parametrized by supercoordinates $Z^{\underline M}=(X^{\underline m},\Theta^{\underline\mu})$.
\begin{eqnarray}
\label{H3} H_3 &=& db_2-\sqrt{T}{A}_3 = \frac 1 {3!}d\xi^p\wedge d\xi^n\wedge d\xi^m H_{mnp} \;  \qquad
\end{eqnarray}
is the generalized field strength of a {\it dimensionless} 2-form gauge potential $b_2= \frac 1 2 d\xi^n\wedge d\xi^m b_{mn}(\xi)$ which is the worldvolume field of the M5-brane,
\begin{eqnarray}
\label{tH=}
\tilde{H}^{mn} & = & *H^{mnr} v_r = \frac 1 {3!\sqrt{-g}}
\epsilon^{mnrlpq} H_{lpq}v_r \; , \qquad \end{eqnarray}
and \begin{eqnarray}\label{v:=} v &= &
d\xi^m v_m = \frac {da(\xi)} {\sqrt{ \partial_n a \, g^{nl}\, \partial_l a}}\; , \qquad v_mv^m=1
\; ,
\end{eqnarray}
where $a(\xi)$ is an auxiliary scalar field which is subject to the restriction  $\partial_m a(\xi)\partial^m a(\xi)\not =0$. The unit vector $v_m$ is chosen to be time-like. This is convenient for further passage to the Hamiltonian formulation, but the choice of the space-like vector $v_m = \frac {\partial_m a(\xi) }{ \sqrt{- \partial a \, g\, \partial a}}$ is equally possible \cite{Perry:1996mk,Pasti:1997gx,Bandos:1997ui,Aganagic:1997zq} in topologically admissible cases \cite{Bandos:2014bva}.

The metric $g_{mn}$ which enters the first and the second term of the action is the induced metric on the worldvolume $\mathcal W^6$
\begin{eqnarray}
\label{g=ind}
g_{mn}(\xi)= {E}_m^{\underline a} \eta_{\underline{ab}}{E}_{n}^{\underline b} ={E}_m^{\underline a} {E}_{n\underline a}.
\end{eqnarray}
${E}_m^{\underline a}$ in (\ref{g=ind}) are the components of the worldvolume pull-back ${E}^{\underline a} =E^{\underline a}({Z}(\xi))=d\xi^m {E}^{\underline a}_m$ of the bosonic supervielbein one-form $E^{\underline a} = dZ^{\underline M}E_{\underline M}^{\underline a}(Z)$
of the 11D supergravity superspace. $E^{\underline a}(Z)$ and its fermionic counterpart $E^{\underline \alpha}$ obey a set of supergravity constraints, the most important of which is
\begin{eqnarray}
\label{Ta=}
T^{\underline a} =DE^{\underline a} =dE^{\underline a}+E^{\underline b}\omega_{\underline b}{}^{\underline a}= -i E^{\underline\alpha} \wedge E^{\underline\beta} \Gamma^{\underline a}_{\underline\alpha\underline\beta}\; ,
\end{eqnarray}
where $\omega_{\underline b}{}^{\underline a}(Z)$ is an $SO(1,10)$ spin-connection one-form and $\Gamma^{\underline a}_{\underline\alpha\underline\beta}$ are $11D$ gamma-matrices in the Majorana representation.

The inverse of the induced metric \eqref{g=ind} is denoted by $g^{mn}(\xi)$,
\begin{eqnarray}
\label{g-1=ind}
g^{mk}(\xi)g_{kn}(\xi)= \delta^m_n . \qquad
\end{eqnarray}
It is used to raise the worldvolume vector indices, e.g.
\begin{eqnarray}
\label{E=gE}
E^{m{\underline a}}(\xi)=g^{mn}(\xi)E^{\underline a}_{n}(\xi) ,  \qquad v^m=g^{mn}v_n\; , \qquad \text{etc.}
\end{eqnarray}
For further analysis it is useful to introduce the vector
\begin{equation}\label{sm:=}
s_m \equiv \frac {\sqrt{-g}}{8} \epsilon_{mnklpq} \tilde{H}^{nk}\tilde{H}^{lp}v^q\; ,
\end{equation}
which, by construction, is orthogonal to $v_m$ and hence space-like,
\begin{equation}\label{tv=0}
s_m v^m=0\; .
\end{equation}
Then we can write the integrand of the first term of the M5-brane action \eqref{SM5=} as follows
\begin{eqnarray}\label{-det(I+F)=}
&T\,\sqrt{-\det\left(g_{mn} + \frac{1}{\sqrt{T}} \tilde{H}_{mn}\right)}
= T\,\sqrt{-g} \left(1 + \frac 1{2T} tr\tilde{H}^2 - {\frac 1{4 T^2}} tr\tilde{H}^4
+ {\frac 1 {8T^2}} (\tr\tilde{H}^2)^2\right)^{1/2}
&
\nonumber \\
&= \sqrt{-g} \, \sqrt{T^2 + \frac T {2} tr\tilde{H}^2 -s^ms_m}\;.&
\end{eqnarray}
The action is invariant under the local $\kappa$-symmetry transformations of the coordinate functions $Z^{\underline M}(\xi)$ which have the following form
\begin{eqnarray}\label{kappaM5}
&& i_\kappa E^{\underline a} := \delta_kZ^{\underline M}  E_{\underline M} ^{\underline a} =0  \; , \qquad \nonumber
\\ && i_\kappa E^{\underline\alpha} := \delta_kZ^{\underline M}  E_{\underline M} ^{\underline\alpha} =: \kappa^{\underline\alpha}(\xi) \; , \qquad \nonumber
\\ && \delta_\kappa b_2=\sqrt{T}\, i_\kappa A_3\; , \qquad
\end{eqnarray}
where the parameter $\kappa^\alpha$ satisfies the condition
\begin{equation}
 \kappa^{\underline \alpha} = - {{\mathbf{\Gamma}}^{\underline\alpha}{}_{\underline\beta}  \kappa^{\underline\beta}}
\label{kap=-bGkap}
\end{equation}
with $\tilde{\mathbf{\Gamma}}$ being the $\kappa$-symmetry projector matrix
\begin{eqnarray}\label{Gkappa=}
& \frac {\sqrt{-\det\left(g_{mn} + \frac{1}{\sqrt{T}} \tilde{H}_{mn}\right)}}
{\sqrt{-g}} \; {\mathbf{\Gamma}} = {\Gamma}^{(6)} + \frac{1}{2\sqrt{T}} \tilde{H}_{mn} ({\Gamma}{}^{mnl}C)v_l
- \frac {1} {T} s_{m}{\Gamma}^{mn}v_n, \qquad {\mathbf\Gamma}^2=1\,, \qquad  \\ \nonumber \\ \label{tG6=} & {} {\Gamma}^{(6)}:= \frac 1{6!} \epsilon_{m_1\ldots m_{6}} {\Gamma}{}^{m_1}\tilde{\Gamma}{}^{m_2}{\Gamma}{}^{m_3}\tilde{\Gamma}{}^{m_4}{\Gamma}{}^{m_5}\tilde{\Gamma}{}^{m_{6}}\; ,\qquad \\ \nonumber \end{eqnarray}
In these expressions
\begin{eqnarray} \label{Gm=EG}
& \Gamma_{m}=E_m^{\underline a}\Gamma_{\underline a} \; , \qquad \Gamma^{m}= g^{mn} \Gamma_n\; , \qquad
\end{eqnarray}
where
\begin{eqnarray}
&
\tilde{\Gamma}_{\underline a}=(\gamma_{\underline a}C)=\gamma_{{\underline a} \alpha}{}^\gamma C_{\gamma\beta}=\tilde{\Gamma}_{{\underline a} \alpha\beta}=\tilde{\Gamma}_{{\underline a} \beta \alpha}\; , \qquad  \\ &
{\Gamma}_{\underline a}\,=(C^{-1}\gamma_{\underline a})= C^{ \alpha\gamma} \gamma_{{\underline a} \gamma}{}^\beta={\Gamma}_{\underline a} ^{\alpha\beta}={\Gamma}_{\underline a} ^{\beta\alpha}\; , \qquad
\end{eqnarray}
$\gamma_{{\underline a} \alpha}{}^\beta $ and $C$ are eleven dimensional gamma matrices and a charge conjugation matrix which are all imaginary in our mostly minus metric convention. In the equations with implicit spinorial indices we do not distinguish charge congugation matrix $C_{\alpha\beta}=-C_{\beta\alpha}$ and its inverse
$C^{\beta\alpha}=-C^{\alpha\beta}$.
Finally,
\begin{eqnarray}
& {\Gamma}^{mn}={\Gamma}^{[m}\tilde{\Gamma}{}^{n]}=\frac 1  2 ({\Gamma}^{m}\tilde{\Gamma}^{n}-{\Gamma}^{n}\tilde{\Gamma}^{m})\; , \qquad \tilde{\Gamma}^{mn}=({\Gamma}^{nm})^T\; , \qquad \nonumber\\ & {\Gamma}^{mnl}={\Gamma}^{[m}\tilde{\Gamma}{}^{n}{\Gamma}^{l]}=\frac 1  3 ({\Gamma}^{m}\tilde{\Gamma}^{nl}+{\Gamma}^{n}\tilde{\Gamma}^{lm} +{\Gamma}^{l}\tilde{\Gamma}^{mn})\; , \qquad \text{etc.}
\end{eqnarray}

\section{Tensionless limit of the M5-brane action}\label{nulM5}

Due to the form of the Dirac-Born-Infeld-like term \eqref{-det(I+F)=}, the zero-tension limit $T\mapsto 0$ of the M5-brane action is straightforward, namely
\begin{eqnarray} \label{SM5-T0=}
S_{M5} \vert_{_{T\mapsto 0}} &=&  -  \int_{\mathcal W^6} d^6\xi \sqrt{-g}\sqrt{-s^ms_m}
-
\frac 1 2 \int_{\mathcal W^6} \, v \wedge H_3 \wedge i_vH_3   \qquad \nonumber \\
 &=&  -  \int_{\mathcal W^6} d^6\xi \sqrt{-g}\sqrt{-s^ms_m}
-
\frac 1 4
\int_{\mathcal W^6} \, d^6\xi \sqrt{-g}\tilde{H}^{mn} H_{mnp}v^p  \; . \qquad
\end{eqnarray}
Notice that in this limit the Wess--Zumino term vanishes and $H_3=db_2$ does not have the contribution of the pull-back of the
11D gauge field $A_3$. Note also that, in contrast to conventional tensionless (null) p-branes \cite{Schild:1976vq,Karlhede:1986wb,Lindstrom:1990qb,Bandos:1990mw,Bandos:1993ma,Bandos:1993pa,Bandos:2006af,Bandos:2014lja}  the (induced) metric \eqref{g=ind} on the $6d$ worldvolume of this tensionless M5-brane remains non-degenerate.
\footnote{By the time this manuscript has been finalized, we are not aware of any consistent alternative tensionless limit procedures which would make the $6d$ worldvolume metric of the M5 brane degenerate. This also concerns the tensionless limit of the $D3$-brane in type IIB supergravity discussed in Appendix \ref{tensionlessD3}.}  As has been already discussed in \cite{Gibbons:2000ck,Townsend:2019ils} in the case in which the worldvolume fields $X^{\underline m}(\xi)$ and $\Theta^{\underline \alpha}(\xi)$ vanish, the limit under consideration can be viewed as a strong limit of the worldvolume gauge field strength $H_3$ in comparison with the $M5$-brane tension. In this limit the $(-s_ms^m)$ term dominates over the T-dependent terms in the DBI-like square root (2.11). Thus the square root of $(-s_ms^m)$ remains in the action (3.1) together with the quadratic term in $H_3$, both having the same conformal weight.

The action clearly maintains the invariance under worldvolume diffeomorphisms and bulk supersymmetry. It also maintains $\kappa$--symmetry \eqref{kappaM5} which reduces to
\begin{eqnarray}\label{kappa=def}
&& i_\kappa E^a := \delta_kZ^{\underline M}   E_{\underline M} ^a =0  \; , \qquad \nonumber
\\ && i_\kappa E^\alpha := \delta_kZ^{\underline M}  E_{\underline M} ^\alpha =: \kappa^\alpha \; , \qquad \nonumber
\\ && \delta_\kappa b_2= 0 \; , \qquad
\nonumber \\
&& \kappa^\alpha = - {\mathbf{\Gamma}}^{\alpha}{}_{\beta}  \kappa^\beta \; \label{kap==-bGkap}
\end{eqnarray}
with
\begin{eqnarray}\label{Gkappa-T0=}
 {\mathbf{\Gamma}} \vert_{T\mapsto 0}=
- \frac {1 } { \sqrt{-s^ls_l}} s_{m}{\Gamma}{}^{mn}v_n \equiv - \frac 1 {\sqrt{-s^ls_l}} (s{\Gamma}) (v\tilde{\Gamma}) \;, \qquad
\mathbf{\Gamma}{}^{2}=1\,.
\end{eqnarray}
A peculiarity of the $\kappa$--symmetry transformations (\ref{kappa=def}) is that the field $b_2$ is not transformed.

Note also that  $\sqrt{-s^ls_l}$ appears in the denominator of the $\kappa$--symmetry projector, which suggests that in general
\be\label{t2=not0}
\sqrt{-s^ls_l}\not =0\,.
\ee
For the $6\times 6$ anti-symmetric matrix $\tilde{H}^{mn}$ (\ref{tH=}), which by construction obeys $\tilde{H}^{mn}v_n=0$ and hence has no more than 2 `eigenvalues' (say $h_+$ and $h_-$), the condition (\ref{t2=not0}) implies that both of these are non-vanishing ($h_+\not=0$ and $h_{-} \not=0$, because
$s^m\propto h_+h_-$ and $-s^ms_m\propto (h_+h_-)^2$).
This is what we mean by stating that the tensionless limit requires the gauge field strength of the worldvolume gauge field of the M5-brane be non-vanishing and non-degenerate.

Similarly to the original M5-brane action \eqref{SM5=}, the tensionless action \eqref{SM5-T0=} is invariant under two PST gauge symmetries. The first PST symmetry acts on the worldvolume fields with a scalar parameter $\varphi(\xi)$ as follows
\begin{eqnarray}\label{PST1==}
\delta_\varphi b_2=  \varphi(\xi)\;  \frac {(i_vH_3+{\cal V}_2)}{\sqrt{\partial a\partial a}}  \; , \qquad \delta_\varphi a(\xi)=\varphi(\xi) \; , \qquad \delta_\varphi Z^{\underline M} (\xi)=0\; , \qquad
\end{eqnarray}
where
\begin{eqnarray}\label{cV2=}
 {\cal V}_2 = \frac 1 2 d\xi^n \wedge d\xi^m  {\cal V}_{mn}\; , \qquad
  {\cal V}_{mn}= \, - \, \frac {\partial \sqrt{-s^ls_l}} {\partial \tilde{H}^{mn}}= \, \frac  {\sqrt{-g}} { 2\sqrt{-s^ss_s} } \epsilon_{mnrlpq} s^{r}v^l \tilde{H}^{pq}\;.
\end{eqnarray}
It ensures the pure auxiliary nature of the scalar field $a(\xi)$ (which, e.g. can be gauge-fixed to coincide with the worldvolume time coordinate $a(\xi)=\xi^0$).

The second PST symmetry acts on the gauge field $b_2$ only
\begin{eqnarray}\label{PST2==}
\delta_{\phi_1} b_2= da\wedge \phi_1\; , \qquad \delta_{\phi_1} a(\xi)=0\; , \qquad \delta_{\phi_1} Z^{\underline M} (\xi)=0\; , \qquad
\end{eqnarray}
with an arbitrary 1-form parameter $\phi_1=d\xi^m \phi_m(\xi)$.

The equations of motion of the bosonic and fermionic embedding coordinates $X^{\underline m}(\xi)$ and $\Theta^{\underline\alpha}(\xi)$ are
\begin{eqnarray}
\label{bEq=}
D_m (T^{mn}{E}_{n\underline{a}})=0\; , \qquad
\\ \label{fEq=}
{ T}^{mn}{E}_{n\underline{a}}\, \Gamma^{\underline{a}}{}_{\underline{\alpha}\underline{\beta}}\,  {E}_{m}{}^{\underline{\beta}}   =0\; , \qquad
\end{eqnarray}
where $D_m=\partial_m+\omega_m$
is a covariant derivative with $\omega^{\underline a}{}_{\underline b}$ being the worldvolume pullback of an $SO(1,10)$ spin connection and  $T^{mn}$ has the following form
\begin{eqnarray}
\label{cTmn=}
{T}^{mn}
&=&  -\, \sqrt{-g}\,\sqrt{-s^ls_l}\,(v^m -\, \hat s^m)(v^n  -\, \hat s^n)\;,\qquad \hat{s}^m=s^m/ \sqrt{-s^ls_l}\,.
\end{eqnarray}
The covariant derivative $D_m$ does not contain the Christoffel symbol associated with the induced worldvolume metric because of the definition of the quantity $T^{mn}$ in \eqref{cTmn=}. Namely, the direct derivation of the  bosonic equation of motion gives $\hat D_m(\frac{1}{\sqrt{-g}}T^{mn}{E}_{n\underline{a}})=0$, where $\hat D_m $ contains the worldvolume Christoffel connection. But, with the use of a well known identity, this equation reduces to \eqref{bEq=}.

In the static gauge $X^m(\xi)=\xi^m$ (fixing the worldvolume diffeomorphisms) the quantity $T^{mn}$ becomes the worldvolume Hilbert energy-momentum tensor of the tensionless M5-brane (as we show in the Appendix \ref{emtofbrane}).

The equation of motion of $b_2$ is
\begin{equation}\label{eomb2}
 da \wedge d \left(\frac {i_vH_3+{\cal V}_2}{\sqrt{\partial a\partial a}}\right)=0.
\end{equation}
It is easily integrated and, with the use of the symmetry (\ref{PST2==}), produces the non-linear self-duality condition of \cite{Townsend:2019ils}
\begin{eqnarray}\label{ivH3=-cV2}
i_vH_3+{\cal V}_2=0 \qquad \Leftrightarrow \qquad H_{mnk}v^k=-{\cal V}_{mn}\; ,
\end{eqnarray}
where ${\cal V}_2$ was defined in (\ref{cV2=}). In Appendix \ref{A} we give an alternative derivation of the non-linear self-duality condition.

\section{ 11D super-Weyl invariance of the tensionless M5-brane action and conformal symmetries of its bosonic sector}\label{confsym}

Since the tensionless M5 action does not contain dimensionful parameters, in addition to the symmetries mentioned in the previous section the tensionless M5-brane action \eqref{SM5-T0=} is also invariant under $11D$ super-Weyl transformations, which also leave invariant the supergravity constrains. These transformations (with a superfield parameter $W(Z)$) act as follows on the supervielbeins
\bea
E^{\underline{a}}(Z) &\mapsto & e^{2W(Z)} E^{\underline{a}}(Z)\,, \nonumber \\
E^{\underline{\alpha}}(Z) &\mapsto & e^{ W(Z)} \left(E^{\underline{\alpha}}(Z) -i  E^{\underline{a} }\Gamma_{\underline{a}}{}^{\underline{\alpha}\underline{\beta}}D_{\underline{\beta}}W(Z) \right)
\eea
and on the superspace spin connection
\be w^{\underline{a}\underline{b}}\mapsto w^{\underline{a}\underline{b}}+4E^{[\underline{a}}D^{\underline{b}]}W+ 2i E^{\underline{c}}DW\Gamma_{\underline{c}}{}^{\underline{a}\underline{b}}DW+2E^{\underline{\alpha}}
\Gamma^{\underline{a}\underline{b}}{}_{\underline{\alpha}}{}^{\underline{\beta}} D_{\underline{\beta}}W\; . \nonumber \ee

In contrast to the case of the fully-fledged super-M5-brane, in the tensionless limit the background supergravity superfields enter the action only through the induced metric \eqref{g=ind} which is transformed  by the super-Weyl rescaling as follows
\be
g_{mn}\mapsto g_{mn}  e^{4W(Z(\xi))}\;.
\ee
As such, the proof of the super-Weyl invariance of the tensionless M5-brane action is straightforward. This implies, in particular, that the tensionless M5-brane action in flat superspace is equivalent to the action in any conformally flat superspace.

It is natural to expect that in flat 11D superspace
for which
\be\label{flatss}
E^{\underline a}=dX^{\underline a}-id\Theta\Gamma^{\underline a}\Theta\,, \qquad E^{\underline \alpha}=d\Theta^{\underline  \alpha}\,.
\ee
the action \eqref{SM5-T0=} is invariant under the 11D spacetime conformal symmetry \footnote{One might also be tempted to search for invariance under a superconformal symmetry, but in conventional  $D>6$ superspaces, without additional exotic coordinates, superconformal supergroups are not known.}.
Let us begin by checking whether the action \eqref{SM5-T0=} is invariant under the rescaling of target-superspace coordinates
\be\label{scaling}
X^{\underline a}\mapsto \lambda X^{\underline a}\; , \qquad \Theta^{\underline \alpha}\mapsto  \sqrt{\lambda}  \Theta^{\underline \alpha} \; .
\ee
By definition this rescaling leaves invariant the worldvolume coordinates
$\xi^m$, the auxiliary scalar $a(\xi)$ and the worldvolume gauge field $b_2$,
\be \xi^m \mapsto \xi^m\; , \qquad a(\xi)\mapsto a(\xi)\, , \qquad b_2\mapsto b_2\; , \qquad H_{mnk}\mapsto H_{mnk}\; . \qquad \ee
This implies  that
\bea
E_m^{\underline a} &\mapsto & \lambda E_m^{\underline a} \; , \qquad \\ \label{gmn-L2gmn}
g_{mn} &\mapsto & \lambda^2 g_{mn}\; , \qquad \\
&& \Rightarrow \qquad g^{mn} \mapsto  \lambda^{-2} g^{mn}\; , \qquad  \sqrt{-g} \mapsto  \lambda^{6} \sqrt{-g} \; , \qquad
\eea
and
\bea
v_m &\mapsto & \lambda v_m\; , \qquad \\
v^m &\mapsto & \lambda^{-1}\, v^m\; , \qquad \\
*H^{mnk}  &\mapsto & \lambda^{-6} *H^{mnk} \; , \qquad \\ \tilde{H}^{mn}  &\mapsto& \lambda^{-5} \tilde{H}^{mn} \; , \qquad \\
s_{m}  &\mapsto & \lambda^{-5} s_{m}\; , \qquad \\
s^{m}s_m  &\mapsto & \lambda^{-12} s^{m}s_m\; , \qquad \\
\sqrt{-s^{m}s_m}  &\mapsto & \lambda^{-6} \sqrt{-s^{m}s_m}\; . \qquad \\ \nonumber
\eea
Using these scaling rules we easily find that both the first and the second term in \eqref{SM5-T0=} are scale invariant.
Notice that all the nontrivial transformations under spacetime scaling symmetry come from the  nontrivial transformations of the induced worldvolume metric $g_{mn}$. This means that the scaling invariance of the tensionless M5 action can be (formally) viewed as the invariance under an effective Weyl symmetry acting on the induced metric (similar to the super-Weyl transformations considered above).

In the bosonic case, in which $\Theta^{\underline \alpha}=0$ and $E^{\underline a}=dX^{\underline a}$ all the conformal symmetry transformations can be obtained from scaling and inversion. The latter acts on the bosonic coordinates of the flat spacetime as follows
\be\label{invX}
X^{\underline a}\mapsto \frac {X^{\underline a}}{X^{\underline b}X_{\underline b}}\; ,
\ee
and results in

\be\label{inv=dXdX}
\partial_m X^{\underline a}\partial_nX_{\underline a} \mapsto \frac 1 {(X^{\underline b}X_{\underline b})^2}\, \partial_mX^{\underline a}\partial_nX_{\underline a}\; .
\ee
Therefore, we find that the induced metric gets rescaled under inversion as follows
\be\label{inv=gind}
g_{mn}\vert_{_{\Theta=0}} \mapsto \frac 1 {(X^{\underline b}X_{\underline b})^2}\,  g_{mn}\vert_{_{\Theta=0}}\,.
\ee
Then the invariance under the inversion of the purely bosonic tensionless 5-brane action is proved in the same way as its scaling invariance. This implies the invariance of the bosonic theory under special conformal symmetry acting on the bosonic coordinate functions as follows
\be\label{sconf=X}
X^{\underline{a}}\mapsto \frac {X^{\underline{a}}+ b^{\underline{a}}X^{\underline{b}}X_{\underline{b}}}{1+ 2 b^{\underline{c}}X_{\underline{c}} + b^{\underline{c}}b_{\underline{c}}\; X^{\underline{e}}X_{\underline{e}}}\; .
\ee
It is instructive also to check this explicitly, at least for the infinitesimal transformations
\be\label{vX=sconf}
\delta X^{\underline{a}}= b^{\underline{a}} (X^{\underline b}X_{\underline b}) -2X^{\underline{a}}\; b_{\underline{b}}X^{\underline{b}}\; .
\ee
Under the conformal boosts the worldvolume derivative of the coordinate functions
transforms by induced $SO(1,9)$ rotations and induced dilatation
\be\label{vdX=sconf}
\delta (\partial_m X^{\underline{a}})=
2 \partial_m  X^{\underline{b}} ( X_{\underline{b}} b^{\underline{a}} -b_{\underline{b}} X^{\underline{a}}) -2 \partial_m  X^{\underline{a}}\; (b_{\underline{b}} X^{\underline{b}}),
\ee
so  the transformation of the induced metric is
\be\label{vgind=sconf}
\delta g_{mn} = - 4 g_{mn}\, (b_{\underline{b}} X^{\underline{b}}) \;.
\ee
Then the proof of the invariance of the bosonic action under special conformal transformations is reduced to the proof of its invariance under the re-scaling of the induced metric.

Thus the purely bosonic sector of the tensionless M5-brane action possesses the 11D spacetime conformal symmetry (forming an $so(2,11)$ algebra) whose infinitesimal transformations act on the coordinate functions $X^{\underline a}(\xi)$ as follows
\be\label{vX=confG}
\delta X^{\underline{a}}= a^{\underline{a}} +  X^{\underline{b}}
L_{\underline{b}}{}^{\underline{a}}+\lambda X^{\underline{a}}+ b^{\underline{a}} X^{{\underline b}}X_{\underline b} -2X^{\underline{a}}\; b_{\underline{b}}X^{\underline{b}}\; \,,
\ee
where $a^{\underline a}$ are the parameters of Poincarè translations, $L_{\underline{b}}{}^{\underline{a}}$ are Lorentz rotations, $\lambda$ is dilatation and $b_{\underline{a}}$ are conformal boosts.

When the coordinate functions orthogonal to the brane worldvolume are zero ($X^i(\xi)=0$, $i=6,\ldots, 10$), this symmetry certainly implies $6d$ conformal invariance of the bosonic theory in the static worldvolume diffeomophism gauge
\be\label{sgauge}
X^m(\xi)=\xi^m\,,\qquad  m=0,1,2,3,4,5\,.
\ee
This is the case  discussed in \cite{Gibbons:2001gy,Townsend:2019ils} in which the theory reduces to the $6d$ conformal non-linear chiral 2-form electrodynamics of a Bialynicki-Birula type \cite{BialynickiBirula:1984tx,BialynickiBirula:1992qj}.

We will now show that, in the static gauge \eqref{sgauge} the $6d$ conformal symmetry is also present when the coordinate functions $X^i(\xi)$ are non-zero, though conformal transformations of the worlvolume coordinates $\xi^m$ get modified and become $X^i(\xi)$-field dependent.

The property of the theory that indicates the presence of conformal invariance is the tracelessness of the energy-momentum tensor. In the flat target space and in the static gauge \eqref{sgauge} the induced worldvolume metric takes the form
\be\label{sinduce}
g_{mn}=\eta_{mn}-\partial_m X^i\partial_n X^i\,,
\ee
and
the energy momentum tensor \eqref{cTmn=} is conserved with respect to the worldvolume coordinates associated with the Minkowski metric $\eta_{mn}$
$$
\partial_mT^{mn}=0,
$$
which follows from the equations of motion \eqref{bEq=}. However,  $T^{mn}$ is traceless with respect to the induced metric but not with respect to $\eta_{mn}$, namely
\be\label{Teta}
T^{mn}g_{mn}=0\quad \rightarrow \quad T^{mn}\eta_{mn}=T^{mn}\partial_m X^i\partial_n X^i= \frac 12 \partial_{m}\partial_n(T^{mn} X^2),
\ee
where
$$
X^2\equiv X^iX^i,
$$
and the right-hand-side was obtained using the field  $X^i(\xi)$ equation of motion
$$\partial_m(T^{mn}\partial_nX^i)=0\,,$$ and the conservation of $T^{mn}$.

Using the form of $T^{mn}\eta_{mn}$ in \eqref{Teta} we get the improved tensor, which is \linebreak $\eta_{mn}$-traceless, in the following form
\bea\label{step2}
\hat T^{mn}&=& T^{mn}+\frac 1{2(d-2)}\Big(2\partial^{(m} \partial_l(T^{n)l} X^2)-\eta^{mn}\partial_p\partial_q(T^{pq} X^2)\\
&&- \partial^{2}(T^{mn}X^2)+\frac 1{(d-1)}[\eta^{mn}\partial^2(T_p{}^p X^2)-\partial^{m}\partial^n(T_p{}^p X^2)]\Big)
\nonumber\\
&=& T^{mn}+\Delta^{mn}\,,\nonumber
\eea
where $d$ is the dimension of the worldvolume.
The tensor is traceless $\hat T_m{}^m$=0, conserved $\partial_m \hat T^{mn}=0$ and symmetric.

Note that the ``improving" term $\Delta^{mn}$ is an identically conserved quantity, $\partial_m\Delta^{mn}\equiv 0$. Hence it can be written as
$$
\Delta^{mn}=\partial_l K^{[lm]n}(\xi)
$$
with $K^{[lm]n}(\xi)$ being a tensor with antisymmetric indices $[lm]$. So the improvement is of the conventional Belinfante form. In particular, $\hat T^{mn}$ and $T^{mn}$ share the same total conserved $d$-momentum $P^m=\int T^{0m}(\xi)\,d^{d-1}\xi$.

The existence of a traceless energy-momentum tensor is a manifestation of the presence of the scale-invariance of the tensionless $M5$-brane which we proved above.
According to a general theorem (see e.g. \cite{Qualls:2015qjb} for a review) the existence of the (improved) conserved symmetric traceless tensor of the form similar to \eqref{step2} means that the scale invariant theory is invariant under the whole conformal symmetry.
In our case this is a $6d$ worldvolume conformal symmetry which has its roots in the $11D$ target-space conformal invariance \eqref{vX=confG}. In particular, upon imposing the static gauge \eqref{sgauge} and restricting the consideration to rescaling and conformal boosts along worldvolume directions we find that \eqref{vX=confG} acts on the worldvolume coordinates $\xi^m$ as follows
\be\label{dcxi}
\delta \xi^m=\lambda \xi^{m}+ b^{m} (\xi_n\xi^n)-2\; b_{n}\xi^{n}\xi^{m}-b^{m} X^iX^i(\xi)\,,
\ee
where the last term is a non-standard field-dependent boost of $\xi^m$. The above transformations are associated with compensating worldvolume diffeomorphisms required to keep the static gauge \eqref{sgauge} intact when the conformal transformations \eqref{vX=confG} act on $X^m$.

In the static gauge \eqref{sgauge}, the conformal transformations of the bosonic fields $X^i(\xi)$ $(i=6,\ldots, 10)$ which follow from the variations \eqref{vX=confG} (with $a^{\underline a}=L_{\underline a}{}^{\underline b}=b^i=0$) accompanied by the compensating worldvolume diffeomorphisms of $\xi^m$ are
\be\label{XIconf'}
\delta X^i(\xi)= X'^{i}(\xi')-X^i(\xi)=\lambda X^{i}-2b_{m}\xi^{m}\,X^{i},
\ee
while the corresponding conformal variation of $X^i$ in the same worldvolume point is
\bea\label{XIconf}
{\underline\delta} X^i(\xi)&=& X'^{i}(\xi)-X^i(\xi)=-\delta \xi^m \partial_mX^i+\frac 16 (\partial_m\,\delta \xi^m)X^i+\frac 16 b^m\partial_m(X^jX^j)\, X^i(\xi)\,.
\eea
The first two terms in the r.h.s. of the above formula are the standard conformal transformations of $6d$ scalar fields of a non-canonical conformal (mass) dimension $\Delta=-1$ (the canonical dimension of the $6d$ scalar field is $m^{2}$, i.e. $\Delta=2$), while the last term indicates that the modified conformal boosts in \eqref{dcxi} are non-linearly realized on $X^i(\xi)$.

The gauge field $b_{mn}(\xi)$ varies under the conformal transformations \eqref{dcxi} similar to its variation under the worldvolume diffeomorphisms

\be\label{deltabmn}
{\underline\delta} b_{mn}=b'_{mn}(\xi)-b_{mn}(\xi)=-\delta\xi^l\partial_l b_{mn}+2b_{l[m}\partial_{n]}\delta\xi^l=-\delta\xi^lH_{lmn}+2\partial_{[m}\Big(b_{n]l}\delta\xi^l\Big)\,.
\ee
The last term can be compensated by gauge transformations of the 2--form potential.

One can directly check that the commutators of the variations \eqref{dcxi}-\eqref{deltabmn} obey the $6d$ conformal algebra.  The closure of the algebra is not surprising, taking into account its origin from the conventional conformal symmetry of the $11D$ target space.

It is now instructive to have a look at the form of the Noether currents associated with the $D=11$ conformal symmetry \eqref{vX=confG} and see how, upon imposing the static gauge \eqref{sgauge}, they reduce  to
the Noether currents of the $6d$ conformal symmetry and other remnants of the $11D$ symmetry that become internal from the point of view of the $6d$ worldvolume.

\section{Noether currents of the target-space conformal symmetry and their worldvolume counterparts}

Before imposing the static gauge \eqref{sgauge}, the variation of the bosonic (part of the) action of the tensionless M5-brane  with respect to the $11D$ coordinate functions $\delta X^{\underline a}$ has the following form
\be
\delta S_{M5}\vert_{T=0}= \int \partial_m (\delta X^{\underline{a}}) \, { T}^{mn} \partial_n X_{\underline{a}}\; ,
\ee
where $T^{mn}$ is given in \eqref{cTmn=}.
The Noether currents for $11D$ spacetime conformal symmetry transformations  \eqref{vX=confG}
can be obtained by promoting their parameters to $\xi$--dependent functions ($a^{\underline{a}}(\xi)$, ..., $b^{\underline{a}}(\xi)$) and reading the quantities that multiply the derivatives of these parameters in the variation of the action
\bea
\delta S_{M5}\vert_{T=0}\dot{=}
 \int d^6\xi\Bigg( \partial_m (a^{\underline{a}}) \, T^{mn} \partial_n X_{\underline{a}} + \partial_m (
L_{\underline{b}\underline{a}}) \, X^{[\underline{b}} \,T^{mn} \partial_n X^{\underline{a}]} + \nonumber\\
+ \partial_m (\lambda) \; \frac  1 2 \, T^{mn} \partial_n (\underline{X}^2) + \partial_m (b_{\underline{b}}) (\eta^{\underline{a}\underline{b}}{\underline X}^2 -2X^{\underline{a}}X^{\underline{b}})\, T^{mn} \partial_n X_{\underline{a}} \Bigg)+ \cdots \; ,
\eea
where   $\underline{X}^2=X^{\underline{a}} X_{\underline{a}}$ and $\cdots$ stand for the terms that vanish on the equations of motion.

Thus we get the Noether currents associated, respectively, with the translations $a^{\underline a}$, Lorentz rotations $L_{\underline {ab}}$, rescalings $\lambda$ and conformal boosts $b^{\underline a}$
\bea
{\mathcal J}^{m\underline{a}} &=& T^{mn} \partial_n X^{\underline{a}} \; , \qquad \label{trans}  \\
{\mathcal J}^{m[\underline{a}\underline{b}]} &=& T^{mn} \partial_n X^{[\underline{a}}X^{\underline{b}]}=  {\mathcal J}^{m[\underline{a}}X^{\underline{b}]}\; , \qquad \\
{\mathcal J}^m &=&  \frac  1 2 \, T^{mn} \partial_n (\underline{X}^2) = {\cal J}^m_{\underline{a}}X^{\underline{a}}\; , \qquad  \\
\tilde{{\mathcal J}}{}^{m\underline{a}}&=& (\eta^{\underline{a}\underline{b}}\underline{X}^2 -2X^{\underline{a}}X^{\underline{b}})\,T^{mn} \partial_n X_{\underline{b}} \nonumber \\ &=&  \underline{X}^2 {\cal J}^{m\underline{a}} - 2 X^{\underline{a}} {\cal J}^m
\; . \qquad
\eea
Upon fixing the static gauge \eqref{sgauge} the Noether current \eqref{trans} of the $11D$ translations splits into two parts, one of which is the $6d$ energy-momentum tensor and another one is associated with the constant shifts of the $6d$ scalar functions $X^i(\xi)$
\bea
{\mathcal J}{}^{m\underline{a}} \qquad = \qquad
\begin{cases}
T^{mn} \cr
{\mathcal J}^{mi} =T^{mn} \partial_n X^{i} \; .
\end{cases}
\eea
The dilatation current becomes
\bea
{\mathcal J}{}^m= \frac  1 2 \partial_n (T^{mn} (\xi^2 - X^iX^i))=T^{mn} (\xi_n - {X^i} \partial_n X^i ) =T^{mn} \xi_n - {\mathcal J}^{mi} X^{i}
\; .
\eea
The $11D$ Lorentz-symmetry current splits into
\bea
{\mathcal J}{}^{m, nl}=T^{m[n}\xi^{l]}\; , \\
{\mathcal J}{}^{m, nj}=\frac 1 2 T^{ml}  (\delta_l{}^n  X^{j}- \xi^{n}\partial_l X^j)\; , \\
{\mathcal J}{}^{m, ij}=T^{mn}\partial_n X^{[i}\; X^{j]}={\cal J}{}^{m, [i}X^{j]}\;,
\eea
and the $11D$ conformal boost current splits into that of the $6d$ conformal boost and the one associated with ``internal'' boosts $b^i$
\bea\label{confboostsplit}
{\hat{\mathcal J}}{}^{m\underline{a}} \qquad = \qquad
\begin{cases}
\hat{\mathcal J}^{mn}=(\xi_l\xi^l-X^iX^i) T^{mn} - 2  {\mathcal J}^m\xi^n \cr
\hat{\mathcal J}^{mi}(\xi_l\xi^l-X^jX^j)T^{mn}\partial_n X^i-2 {\mathcal J}^m X^i \; .
\end{cases}
\eea

We have thus revealed peculiar features of the  worldvolume conformal invariance of the tensionless bosonic M5-brane theory in the static gauge, which has its origin in the 11D target-space conformal symmetry of the action. This conformal symmetry is non-linearly realized on the worldvolume scalar fields $X^i(\xi)$ and the chiral two-form $b_2(\xi)$, and involves field dependent transformations of the worldvolume coordinates.

\section{Conclusion and outlook}

In this paper we have elaborated on the tensionless limit of the complete supersymmetric action of the M-theory 5-brane and the peculiar properties of the conformal symmetry of its bosonic sector.

Concerning the complete tensionless M5-brane theory in target superspace (with $\Theta\not= 0$), at the moment it is not clear to us whether it can possess a full superconformal invariance in the case in which  $11D$ target superspace is flat. It is rather unlikely that such superconformal symmetry may originate from the superconformal invariance in $11D$ target superspace. This is related to the fact that the only known $11D$ generalization of the superconformal group is $OSp(1|64)$. It cannot be realized solely on the superspace coordinates $Z^{\underline M} =(X^{\underline m},\Theta^{\underline\mu})$ and requires the introduction of additional bosonic coordinates.\footnote{See \cite{Bandos:2002nn} for an extensive discussion of this issue and further references.} An alternative can be to consider the tensionless M5-brane in an $AdS_7\times S^4$ superspace whose superisometries may effectively act as superconformal transformations on the worldvolume fields realizing a $6d$ $\mathcal N=(4,0)$ superconformal symmetry. In this respect results on the form of the M5-brane action in the $AdS_7\times S^4$ superbackground and issues with gauge fixing its local worldvolume symmetries studied in \cite{Claus:1998mw,Claus:1998fh,Pasti:1998tc} can be useful.

To see whether such superconformal symmetry may arise on the worldvolume of the tensionless M5-brane, one should completely gauge fix worldvolume diffeomorphisms and local fermionic $\kappa$-symmetry. Because of the specific form \eqref{Gkappa-T0=} of the $\kappa$-symmetry projector, finding an appropriate gauge choice has turned out to be a non-trivial technical problem which we will address in a future work. Similar questions regarding the superconformal structure of the tensionless D3-brane can be studied using the action given in Appendix \ref{tensionlessD3}.

\subsection*{Acknowledgements}
The authors are thankful to Paul Townsend for the fruitful collaboration at early stages of this project.
Work of IB and DS was supported in part by the Spanish  MCI, AEI and FEDER (ERDF EU) grant  PID2024-155685NB-C21``Gravity, Supergravity and Superstrings'' (GRASS), as well as by the Basque Government Grants IT1628-22 and IT1977-26. Work of DS was also supported in part by the CARIPARO Foundation under grant No.~68079 and by the Australian Research Council project DP230101629.

\bigskip

\section*{\LARGE{Appendices}}
\appendix

\section{An alternative derivation of the self-duality condition from the PST action}\label{A}
The equation of motion of the $b_2$ gauge field which one directly gets  from the PST action like \eqref{SM5=} or \eqref{SM5-T0=} has the following form
\be\label{free}
  \partial_m \Big(\frac{\sqrt{-g}}2 v^{[m} \tilde{H}{}^{nl]} - \frac 1{12}\varepsilon^{mnlpqr}H_{pq}v_r - \frac 1 6\varepsilon^{mnlpqr}\mathcal V_{p q}v_r\Big)=0\,,
  \ee
  where  $H_{mn}=H_{mnp}v^p$,
    $\mathcal V_{pq}=\frac {\partial \mathcal V(\tilde H)}{\partial \tilde{H}^{pq}}$ \footnote{The derivative with respect to $\tilde H^{pq}$ is defined such that
   $\frac{\partial {\tilde H}^{mn}}{\partial \tilde H^{pq}}=\delta_p^m \delta_q^n - \delta_q^m \delta_p^n$.}  and $\mathcal V(\tilde H)$ is a function of $\tilde{H}^{mn}$ that appears in the PST action.
  For instance,  for the M5-brane
  $$\mathcal V(\tilde H)=-\, T\, \frac 1 {\sqrt{-g}}\, \sqrt{-\det \left(g_{mn}+\frac 1{\sqrt T}\tilde H_{mn}\right)}\,$$
   and for its tensionless limit
   $$\mathcal V(\tilde H)=-\sqrt{-s^ms_m}\,.$$
    In all the previous literature on the PST formulation this equation was solved as follows. We used the identity
  \be\label{identity}
  \star H^{mnl}=3\tilde H^{[mn}v^{l]}+\, \frac 1 {2\sqrt{-g}}\,  \varepsilon^{mnlpqr}H_{pq}v_r
  \ee
  to replace the first term in \eqref{free} with
  \be
  \tilde H^{[mn}v^{l]}=\, \frac 1 3 \star H^{mnl}-\frac {1} {6\sqrt{-g}}\, \varepsilon^{mnlpqr}H_{pq}v_r\; .
\ee
Then, since $\partial_m(\sqrt{-g}\star H^{mnl})\equiv 0$, we get
\be\label{free1}
\frac 16\partial_m [\varepsilon^{mnlpqr}(\mathcal V_{pq}+H_{pqs}v^s)v_r]=0\,.
\ee
It is this equation that we always solved, with the use of one of the PST symmetries, to arrive at
\be H_{mnp}v^p=-\mathcal V_{mn}
\ee
as was reviewed for the tensionless $M5$ in the end of  Section \ref{nulM5}.

Now, what happens if instead we use \eqref{identity} to replace the second term in \eqref{free}. We thus get the equation
\be\label{free0}
  \partial_m \Big(\sqrt{-g}\,v^{[m} \tilde{H}{}^{nl]} - \frac 16\varepsilon^{mnlpqr}\mathcal V_{p    q}(\tilde H)v_r\Big)=0\, .
  \ee
Its solution is
\be\label{sol1}
v^{[m} \tilde{H}{}^{nl]} -\frac 1{6\sqrt{-g} }\varepsilon^{mnlpqr}\mathcal V_{p q}v_r
=\frac 1{6\sqrt{-g} }\varepsilon^{mnlpqr}\partial_pA_{qr}=\frac 1 3 \star F^{mnl},
\ee
where $A_{qr}$ is a two-form different from $b_2$ and $F_3=dA_2$.
The Hodge dual of \eqref{sol1} is
\be\label{Fnl}
F^{mnl}=-\, 3v^{[m} \mathcal V{}^{nl]}(\tilde H) +\frac 1{2\sqrt{-g} }\varepsilon^{mnlpqr}\tilde H_{p q}v_r.
\ee
Note that, since $\tilde H^{mn}$ is inert under the PST symmetry \eqref{PST2==}, the three-form $F_3$ is also invariant under this symmetry.

Comparing eq. \eqref{Fnl} with the identity \eqref{identity} applied to $F_3$ we find that
\be\label{F=H}
i_vF_3=-\mathcal V_2(\tilde H),\qquad i_v\star F_3=\tilde H_2,
\ee
which is similar to \eqref{ivH3=-cV2},
or equivalently
\be\label{Fsd}
i_vF_3=-\mathcal V_2(i_v\star F)\,.
\ee
Eq. \eqref{Fsd} means that $F_3$ is non-linearly self-dual. From the above relations it follows that, even if (in general) there is no a simple relation between $H_3$ and $F_3$, on the mass shell one can forget about $H_3$ and use the non-linearly self-dual $F_3$ as the physical field strength.

For the tensionless M5 Eq. \eqref{free0} reads
\be\label{b2=Eqs}
\partial_m (\sqrt{-g} (v-\hat{s})^{[m} \tilde{H}{}^{nl]} )=0\;
\ee
and  $F_{pqr}= \frac {\sqrt{-g}}{2}\epsilon_{pqrmnl} (v-\hat{s})^{m} \tilde{H}{}^{nl}$.

\section{Worldvolume energy-momentum tensor on a brane}\label{emtofbrane}

A prescription to compute the energy-momentum tensor of a worldvolume field theory on the branes is as follows (see e.g. \cite{Brizio:2026ynw}). For the purpose of this paper we will consider the case of a purely bosonic flat background, but the procedure works also for certain curved backgrounds and in superspace.
\begin{itemize}
\item
Impose the static gauge $\xi^m=X^m$ in the Lagrangian density $\mathcal L$. Then the induced metric becomes $g_{mn}=\eta_{mn}-\partial_m X^i\partial_n X^i$.
\item
Replace $\eta_{mn}$ with an intrinsic (independent) curved metric $\gamma_{mn}(\xi)$.
\item
Compute the Hilbert energy-momentum tensor
\be\label{HT}
T^{mn}=\frac 2{\sqrt{-\det \gamma}}\frac{\delta\mathcal L}{\delta \gamma_{mn}}=\frac 2{\sqrt{-\det \gamma}}\frac{\delta\mathcal L}{\delta g_{mn}}\,.
\ee
\item
In \eqref{HT} replace  $\gamma_{mn}(x)$ back with $\eta_{mn}$. This results in
\be\label{T}
T^{mn}=2\frac{\delta\mathcal L}{\delta g_{mn}}\,.
\ee
\end{itemize}
For the bosonic branes described by the Nambu-Goto Lagrangian density
\be
\label{NGL}\mathcal L_{NG}=\sqrt{-\det g_{mn}}=\sqrt{-\det (\eta_{mn}-\partial_m X^i\partial_m X^i})
\ee
we thus have
$$
T^{mn}=\sqrt{-\det g}\,g^{mn}\,.
$$
It coincides with the canonical form of the energy-momentum tensor of the scalar fields $X^i(\xi)$ computed with the Noether procedure applied to \eqref{NGL}. Therefore this tensor is conserved by construction in the static gauge: $\partial_m T^{mn}=0$. Indeed, the equations of motion of $X^{\underline a}=(X^l,X^i)$ before imposing the static gauge are
\be
\partial_m(\sqrt{-g}g^{mn}\partial_nX^{l})=0\,, \qquad
\partial_m(\sqrt{-g}g^{mn}\partial_nX^i)=0\,, \qquad
\ee
$$l=0,\ldots, p; \qquad i=D-p-1,\ldots, D-1. $$
In the static gauge $\xi^l=X^l$ the first of these equations  reduces to
\be\label{conservation}
\partial_m(\sqrt{-g}g^{mn})=\partial_mT^{mn}=0.
\ee
Let us now apply this procedure to the bosonic sector of the tensionless $M5$ action \eqref{SM5-T0=}.
The variation of its first term with respect to the induced metric results in
\bea
-\delta\left( \sqrt{-g}\sqrt{-s^ms_m}\right)=-\frac 12\sqrt{-g}\sqrt{-s^ls_l}(v^nv^m+\hat{s}^m \hat{s}^n)\delta g_{mn} \; ,
\eea
while the metric variation of the second term in \eqref{SM5-T0=} is
\bea\label{vartHHv=}
-\frac 14 \delta \left(\sqrt{-g}\tilde{H}^{mn} H_{mnp}v^p\right)=
\sqrt{-g}\sqrt{-s^ls_l} v^n\hat{s}^m \delta g_{mn}
\; ,  \qquad
\eea
where to derive the right-hand side we used the identity
\be
\tilde{H}^{mn} H_{mnp}= \tilde{H}^{mn} H_{mnq} v_pv^q+4{s}_p\; .
\ee
Substituting these variations into \eqref{T} we get the energy-momentum tensor of the bosonic tensionless M5 brane
\be
T_{M5}^{mn}|_{{}_{T=0}}=- \sqrt{-g}\sqrt{-s^ls_l}(v-\hat{s})^m (v-\hat{s})^n \; .
\ee
Note, once again, that this tensor is conserved with respect to the flat metric $\eta_{mn}$, like \eqref{conservation}. However, it is traceless with respect to the induced metric $g_{mn}$ and not with respect to the flat worldvolume metric, unless $X^i=const$. The stress-tensor can be improved to be traceless with respect to the metric $\eta_{mn}$ as described in Section \ref{confsym}.

\section{Tensionless D$3$-brane, or ${\cal N}=4,\,\, D=4$ supersymmetric Bialynicki--Birula theory}\label{tensionlessD3}

The action of a tensionless $D3$-brane looks similar to that of \eqref{SM5-T0=}. It can be obtained by taking a tensionless limit of the duality-symmetric action of the $D3$-brane constructed in \cite{Berman:1997iz,Nurmagambetov:1998gp}. In the conventions of \cite{Bandos:2020hgy} the result is
\be\label{T0D3}
S_{D3} \vert_{T\mapsto 0}
 =  \int_{\mathcal W^4} d^4\xi \sqrt{-g}\, \Big(\frac12\varepsilon_{IJ}\, E^{I}\cdot B^{J} -\sqrt{-p^m p_m}\Big)\, \qquad I,J=(1,2)
 ,
\ee
where $E^I_{m}=F_{mn}^Iv^n$, $B^I_{m}=\star F_{mn}^Iv^n$ and $F^I_{mn}=\partial_mA^I_n-\partial_nA^I_m$ are field strengths of a doublet of worldvolume gauge vector fields $A_m^I(\xi)$. The induced metric on the $D3$-brane worldvolume is constructed from the pull-backs of vector supervielbeins of type IIB $D=10$ supergravity
\bea\label{4dmetric}
& g_{mn}=(\partial_{m}Z^{\underline M} E_{\underline M} {}^{a})( \partial_{n}Z^{\underline N} E_{\underline N} {}^{b})\eta_{ab}\,,\qquad Z^{\underline M} =(X^{\underline m},\Theta^{\alpha}_1, \Theta^{\beta}_2) &\\
&\underline m=0,1,\ldots, 9;\qquad \alpha=1,\ldots, 16 &\nonumber
\eea
and
\be\label{smD3}
p^{m}=\frac 12\varepsilon_{IJ}\varepsilon^{mnpq}B^I_{n}B^J_{p}v_q\,.
\ee
The analysis of this theory can be performed in a way similar to the M5-brane case.

\bigskip

\providecommand{\href}[2]{#2}\begingroup\raggedright\endgroup

\end{document}